\documentclass[twoside,prl,twocolumn,showpacs]{revtex4}
\pdfoutput=1
\usepackage{graphicx}
\usepackage{amsmath}
\usepackage{amssymb}
\usepackage{wrapfig}
\usepackage{bbm}
\usepackage[colorlinks,urlcolor=green,citecolor=blue]{hyperref}
\usepackage{bbm}

\newcommand{\abso}[1]{\left\vert #1 \right\vert}
\newcommand{\ket}[1]{\vert #1 \rangle}
\newcommand{\bra}[1]{\langle #1 \vert}

\newcommand{\scalar}[2]{\left\langle\, #1 \,\vert\, #2 \,\right\rangle}
\newcommand{\twoarray}[4]{\left(\begin{array}{cc} #1 & #2 \\[.2cm] #3 & #4 \end{array}\right)}
\newcommand{\twovector}[2]{\left(\begin{array}{c} #1 \\[.2cm] #2 \end{array}\right)}

\begin{document}

\title{Magnetic field induced localization in carbon nanotubes}
\author{Magdalena Marga\'{n}ska}
\author{Miriam del Valle}
\author{Sung Ho Jhang}
\author{Christoph Strunk}
\author{Milena Grifoni}
\affiliation{University of Regensburg, 93 053 Regensburg, Germany}
\date{\today}
\begin{abstract}
The electronic spectra of long carbon nanotubes  (CNTs) can, to a very
good approximation, be obtained using the dispersion relation of
graphene with both angular and axial periodic boundary conditions. In
short CNTs one must account for the presence of open ends, which
may give rise to states localized at the edges.
We show that when a magnetic field is applied parallel to the tube axis, it modifies both momentum quantization
conditions, causing hitherto extended states to localize near the
ends. This localization is gradual and initially the involved states
are still conducting. Beyond a threshold value of the magnetic field, which
depends on the nanotube chirality and length, the localization is
complete and the transport is suppressed.
\end{abstract}
\pacs{73.63.Fg, 75.47.-m, 73.23.Ad, 85.75.-d}
\maketitle

%
%
The existence of geometry-induced localized states at the zigzag edge of graphene nanoribbons has been predicted some years ago \cite{nakada:prb1996,brey:prb2006b}, recently seen experimentally
and shown to influence the transport in graphene quantum dots \cite{ritter:natmat2009}. Similar states have been observed at the ends of a single-wall chiral nanotube studied in \cite{kim:prl1999}. In zigzag-armchair nanotube junctions, the interface states calculated to appear at the junction were identified with the end states of the zigzag nanotube fragment \cite{santos:prb2009}.

In this Letter we predict the occurrence  of localized states in CNTs, which is entirely due to the presence of a parallel magnetic field. Above a threshold flux $\phi_{loc}$ (see Eq. (\ref{eq:philoc-spin})) these states decay exponentially with the distance from the nanotube end. They appear even in those chiral CNTs which have no localized end states when the magnetic field is absent. They can be found by a purely analytical method based on the Dirac equation in graphene, and the resulting energy spectrum is the same as that obtained by the numerical diagonalization of the full nanotube Hamiltonian.

\begin{figure}[h]
\includegraphics[width=0.45\textwidth]{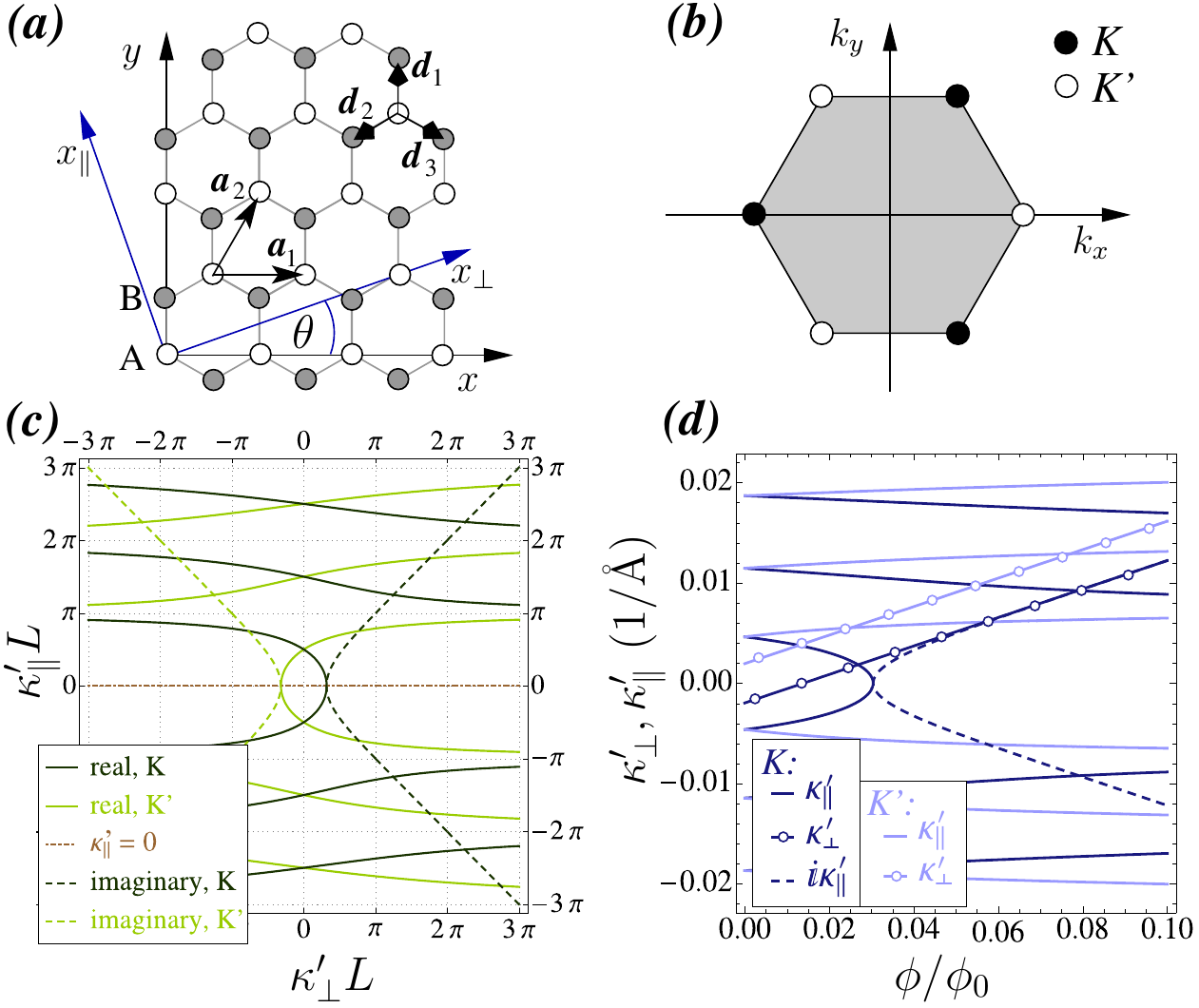}
\caption{\label{fig:general}
$(a)$ Fragment of a graphene lattice. When considering a CNT with chiral angle $\theta$, we shall use the system of coordinates defined by directions perpendicular ($x_\perp$) and parallel ($x_\parallel$) to the tube axis. $(b)$ Brillouin zone of graphene. $(c)$ Real and imaginary solutions of Eq. (\ref{eq:constraint}), determining the quantization of $\kappa_\parallel'$ as a function of $\kappa_\perp'$. $(d)$ Some of the real and imaginary solutions of Eq. (\ref{eq:constraint}) for the Fermi subband ($\kappa_\perp = 0$) of an (18,0) CNT with 100 unit cells, with $\kappa_\perp'$ and $\kappa_\parallel'$ as functions of the magnetic flux $\phi$.}
\end{figure}

{\em The model}. 
Our starting point is the tight-binding Hamiltonian for a honeycomb lattice with one $p_z$ orbital per atom and with the interatomic potential $\hat{V}$. If we calibrate our energy scale so that the on-site energies vanish, 
the Hamiltonian is given by 
\begin{equation}
\label{eq:hamiltonian-real}
\hat{H} = \sum_{i\neq j} t_{ij} \ket{z_j}\bra{z_i}, 
\end{equation}
where $i$ and $j$ are the lattice site indices, $\ket{z_j}$ is a $p_z$ orbital at site $j$ and $t_{ij} = \bra{z_j}\hat{V}\ket{z_i}$ is the hopping integral between the sites. This Hamiltonian nicely captures the properties of flat graphene and CNTs. In order to properly describe finite size nanotubes in magnetic field it is necessary to include the Peierls phase and curvature effects in the hopping elements $t_{ij}$. We follow here the approach of Ando \cite{ando:jpsj2000}.
For the sake of clarity we shall initially neglect the spin-orbit coupling and the Zeeman effect, as they do not change our main conclusion. The spin-dependent effects will be addressed later. \\
The graphene coordinate system and the relevant real space vectors are shown in Fig. \ref{fig:general}(a), while the graphene Brillouin zone with $K$ and $K'$ points is shown in Fig. \ref{fig:general}(b). In order to find the appropriate boundary conditions and eigenstates of CNTs we use an approach based on the Dirac equation treatment \cite{brey:prb2006b,koller:njp2010}.

%
%
{\em Parallel magnetic field}. 
The magnetic field modifies all hopping integrals by a Peierls phase factor. Its form can be derived using the substitution $\mathbf{p} \rightarrow \mathbf{p} - e\mathbf{A}$ and reads
\begin{equation}
t_{ij}(\mathbf{B}) = t_{ij}(0)\;\exp\left\{ \frac{ie}{\hbar} \int_{\mathbf{r}_i}^{\mathbf{r}_j}
  \mathbf{A}(\mathbf{r})\cdot d\mathbf{r}\right\}. 
\end{equation}
In the cylindrical coordinates $(r,\varphi,z)$, with the $z$ direction aligned with the axis of the nanotube, a parallel magnetic field has coordinates $\mathbf{B} = (0,\,0,\,B)$. In the tangential gauge this gives $\mathbf{A} = (0,\, Br/2,\,0)$. The Peierls phase then becomes
\begin{equation}
\frac{ie}{\hbar}\int_{\mathbf{r}_i}^{\mathbf{r}_j} \mathbf{A}(\mathbf{r})\cdot d\mathbf{r} = i \frac{\phi}{\phi_0} (\varphi_j - \varphi_i),
\end{equation}
where $\phi$ is the magnetic flux threading the nanotube, $\phi_0 = h/e$ the flux quantum, 
and $\varphi_j-\varphi_i$ is the difference between the angular coordinates of site $j$ and site $i$.

%
%
{\em Curvature}. In a nanotube the $\sigma$ bonds are not orthogonal to the $p_z$ orbitals and the hopping integral $t_{ij}$ can be expressed as
\begin{equation}
\label{eq:potential}
t_{ij}(0) = \bra{z_j} \hat{V} \ket{z_i} = V_\pi\;\mathbf{n}_{i\text{n}}\cdot\mathbf{n}_{j\text{n}} + V_\sigma\;
\mathbf{n}_{i\text{t}}\cdot\mathbf{n}_{j\text{t}},
\end{equation}
where $V_\pi$ and $V_\sigma$ are hopping parameters for the corresponding bonds \cite{ando:jpsj2000}. In our calculations we shall use the parameters from \cite{bulaev:prb2008}, $V_\sigma = 6.38$ eV and $V_\pi = -2.66$ eV. The vector $\mathbf{n}_{i}$ is a unit vector normal to the nanotube surface at the site $i$. The components $\mathbf{n}_{i\text{n}}$ (normal) and $\mathbf{n}_{i\text{t}}$ (tangential) are defined with respect to a plane containing the $\sigma$ bond between $i$ and $j$ and parallel to the CNT axis. The hopping integral $t_{ij}(0)$ then reads
\begin{eqnarray}
\bra{z_j}\,\hat{V}\,\ket{z_i} & = & V_\pi\,\cos(\varphi_i-\varphi_j)  \nonumber \\
 & - & (V_\sigma-V_\pi)\frac{R^2}{a_c^2}\,\bigl[ 1-\cos(\varphi_i-\varphi_j)\bigr]^2, \label{eq:hopping}
\end{eqnarray}
where $R$ is the nanotube radius and $a_c = 1.42${\AA} is the bond length in graphene. \\
In order to find the CNT spectrum it is convenient to express the Hamiltonian 
 in the Bloch wave basis. The Bloch waves for the CNT sublattice $p$ are given by \cite{saito:1998}
\begin{equation}
\label{eq:bloch-waves}
\ket{\Phi_p(\mathbf{k})} = \frac{1}{\sqrt{N}} \sum_{i=1}^N e^{i\mathbf{k}\cdot\mathbf{r}_{pi}} \ket{z_{pi}}, 
\end{equation}
where $N$ is the number of the unit cells. The Bloch wave on the whole lattice is a linear combination of Bloch waves on individual sublattices 
and can be written as
\begin{equation}
\label{eq:bloch-lattice}
\ket{\Phi(\mathbf{k})} = \sum_{p=A,B} \eta_p (\mathbf{k})\ket{\Phi_p(\mathbf{k})}  = \twovector{\eta_A(\mathbf{k})}{\eta_B(\mathbf{k})}. 
\end{equation}
In this basis the Hamiltonian acquires the form
\begin{equation}
\label{eq:hamiltonian-general}
\hat{H}(\mathbf{k}) = \twoarray{0}{H_{AB}(\mathbf{k})}{H_{AB}^\dag(\mathbf{k})}{0},
\end{equation}
where $H_{AB}(\mathbf{k}) = \sum_{i=1}^3 t_i\,e^{i\mathbf{k}\cdot\mathbf{d}_{i}}$, $\mathbf{d}_{i}$ are the vectors connecting an $A$ sublattice atom with its neighbours, as shown in Fig. \ref{fig:general}(a), and $t_i$'s are the hopping integrals between an atom on sublattice $A$ and its neighbours. Because both the magnetic field and the curvature are uniform along the whole nanotube, the $t_i$'s do not depend on the position of the initial $A$ atom. 
This Hamiltonian can be further expanded around the $K$ ($\tau = 1$) and $K'$ ($\tau = -1$) points (see Fig. \ref{fig:general}(b)), yielding
\begin{equation}
\label{eq:hamiltonian-dirac}
\hat{H}_\tau(\kappa) = \hbar v_F\;\twoarray{0}{e^{i\tau\theta}(\tau \kappa_\perp'+i\kappa_\parallel')}{e^{-i\tau\theta}(\tau\kappa_\perp' -i \kappa_\parallel')}{0}, 
\end{equation}
where $\theta$ is the CNT chiral angle,  indices $(\perp,\parallel)$ denote the components perpendicular and parallel to the nanotube axis respectively, $\hbar v_F = 3\abso{V_\pi} a_c/2$, $\kappa = \mathbf{k} - \tau\mathbf{K}$ and
\begin{subequations}
\label{eq:kappas}
\begin{eqnarray}
\label{eq:kappa-perp}
\kappa_\perp' & = & \kappa_\perp  + \tau\Delta k_\perp^c  + \frac{1}{R}\, \frac{\phi}{\phi_0},\quad \\[.2cm]
\label{eq:kappa-parallel}
\kappa_\parallel' & = & \kappa_\parallel + \tau\Delta k_\parallel^c.
\end{eqnarray}
\end{subequations}
The last term in (\ref{eq:kappa-perp}) is the Aharonov-Bohm contribution while $\Delta k_\perp^c$ and $\Delta k_\parallel^c$ are due to the curvature, 
\begin{subequations}
\label{eq:deltas}
\begin{eqnarray}
\label{eq:delta-kperp}
\Delta k_\perp^c & = & \frac{a_c}{4R^2} \left( 1+ \frac{3}{8} \frac{V_\sigma - V_\pi}{V_\pi}\right)\,\cos(3\theta),\\
\label{eq:delta-kparallel}
\Delta k_\parallel^c & = & -\frac{a_c}{4R^2} \left( 1+ \frac{5}{8} \frac{V_\sigma - V_\pi}{V_\pi}\right)\,\sin(3\theta).
\end{eqnarray}
\end{subequations}
In this derivation we used a small angle approximation, $\sin(\varphi_i-\varphi_j)\simeq \varphi_i-\varphi_j$, which is good already for CNTs with $R\gtrsim 5$\AA. The energy eigenvalues of the Hamiltonian are
\begin{equation} \label{eq:eigenvalues}
E_\pm = \pm \hbar v_F\, \sqrt{(\kappa_\perp')^2 + (\kappa_\parallel')^2}, \quad\quad \tilde{E}_\pm := E_\pm/(\hbar v_F). 
\end{equation}

%
%
{\em Eigenfunctions of the Hamiltonian}. The energy eigenstates are a linear combination of Bloch waves. Since we have expanded the Hamiltonian around the $K$ and $K'$ points, the corresponding Bloch waves and the coefficients $\eta_p$ acquire the index $\tau$. We shall be using
\begin{equation}
\label{eq:psitau}
\Phi_{\tau p}(\mathbf{r},\mathbf{\kappa})=\scalar{\mathbf{r}}{\Phi_p(\tau\mathbf{K}+\kappa)}.
\end{equation}

%
%
{\em Angular boundary condition}. The wave function in the angular direction must be periodic. This imposes
\begin{eqnarray}
 \Phi_{\tau p}((2\pi R,x_\parallel),\mathbf{\kappa}) & \overset{!}{=} & e^{i2\pi n}\;  \Phi_{\tau p}((0,x_\parallel),\mathbf{\kappa}), \nonumber \\
 & \Rightarrow & (\tau K_\perp + \kappa_\perp) = \frac{n}{R}, \label{eq:periodic}
\end{eqnarray}
which is the standard quantization condition \cite{saito:1998}. 

\begin{figure}[h]
\includegraphics[width=0.45\textwidth]{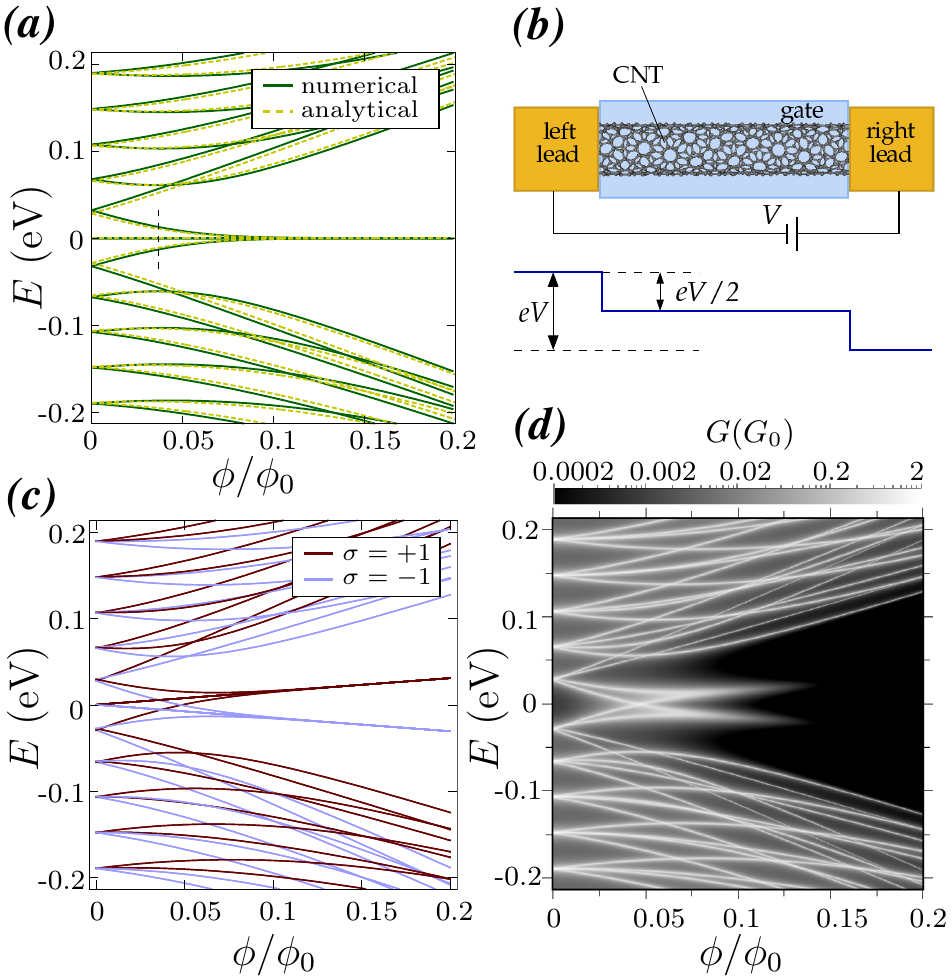}
\caption{\label{fig:18-0} Magnetic field induced localization in a (18,0) zigzag CNT with 100 unit cells ($L = 42.6$ nm, $\theta = 0^\circ$). 
$(a)$ The spectra close to the Fermi level obtained by a numerical diagonalization of the real space Hamiltonian (\ref{eq:hamiltonian-real}) and analytically from the Dirac-like dispersion (\ref{eq:eigenvalues}) with $\kappa_\perp=0$ ($E\neq 0$ states) and $\kappa_\perp = \pm 1/R$ ($E\equiv 0$ states), where $\kappa_\parallel$ is defined by (\ref{eq:constraint}), neglecting the spin. The black dashed line marks the onset of the localization. 
$(b)$ The setup used for the conductance calculation.
$(c)$ Spectra obtained analytically with the electron spin included through (\ref{eq:so-shift}). 
$(d)$ Greyscale plot of conductance, in units of conductance quantum $G_0 = 2e^2/h$, as a function of $\phi$ and the chemical potential $E$, including spin effects.}
\end{figure}

\begin{figure}[ht]
\includegraphics[width=0.45\textwidth]{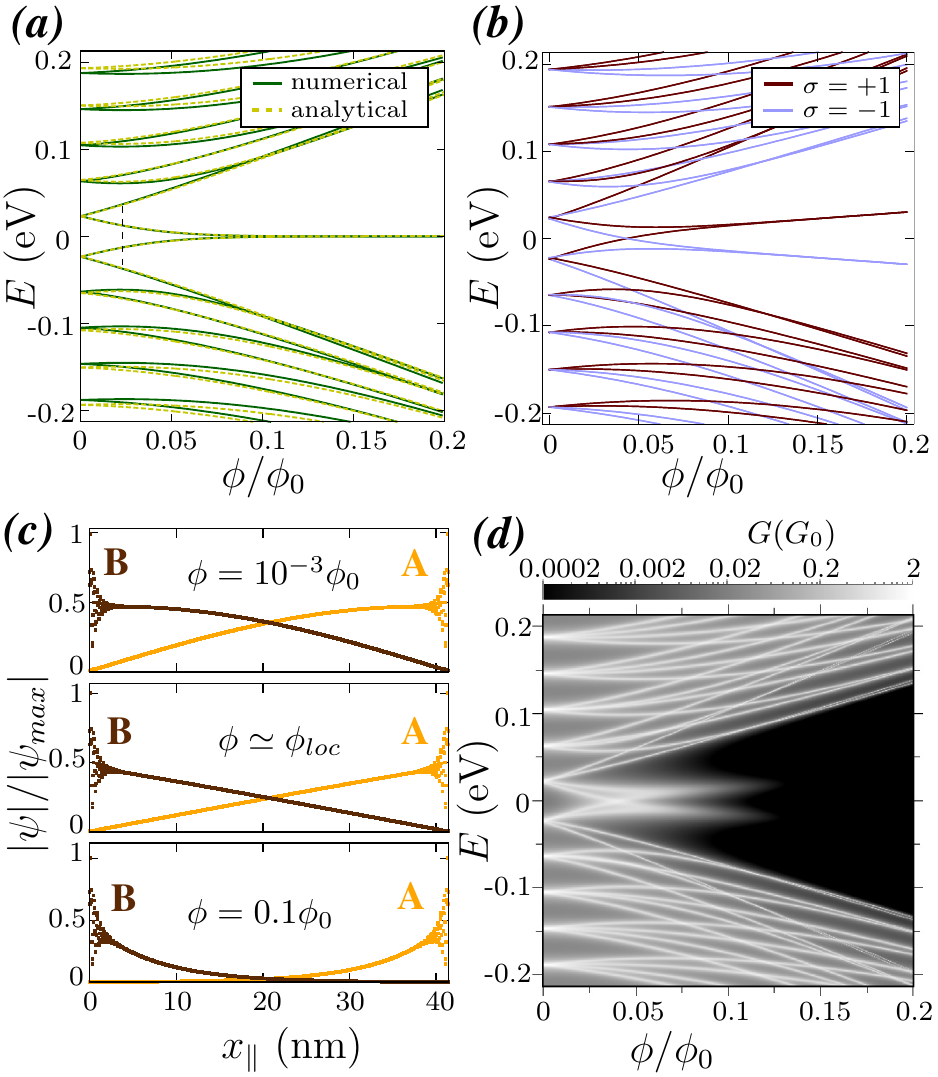}
\caption{\label{fig:12-9} Localization in a (12,9) chiral CNT with 16 unit cells ($L = 41.5$ nm, $\theta = 25^\circ$). 
$(a)$ Comparison of numerical and analytical spectra, neglecting the spin. In this case there are no localized states at $\phi=0$. The dashed line marks the onset of the localization. 
$(b)$ Analytical spectrum including spin effects.
$(c)$ The amplitude of the highest valence eigenstate (obtained numerically) at each atom, projected onto $x_\parallel$, for different values of $\phi$ and neglecting the spin.
$(d)$ Conductance as a function of magnetic flux $\phi$ and chemical potential $E$, including spin effects. }
\end{figure}

%
%
{\em Axial boundary condition}. The wave function at the ends of the nanotube must satisfy open boundary conditions. We shall derive them for a zigzag nanotube ($\theta=0^\circ$), but they are valid for any other chirality except armchair \cite{akhmerov:prb2008}, provided the nanotube edge is a so-called minimal boundary (there are no atoms with only one neighbour). \\
The Hamiltonian (\ref{eq:hamiltonian-dirac}) acting on the wave functions $\ket{\Phi_\tau(\mathbf{\kappa})}$,  (\ref{eq:bloch-lattice}) with (\ref{eq:psitau}), gives two equations:
\begin{eqnarray*}
e^{i\tau\theta} \left[ \tau\kappa_\perp'+i\kappa_\parallel'\right]\,\eta_{\tau B}(\mathbf{\kappa}) & = & \tilde{E}_{\pm} \,\eta_{\tau A} (\mathbf{\kappa})\\[.2cm]
e^{-i\tau\theta} \left[ \tau\kappa_\perp' -i\kappa_\parallel'\right] \,\eta_{\tau A}(\mathbf{\kappa}) & = & \tilde{E}_{\pm} \,\eta_{\tau B}(\mathbf{\kappa}).
\end{eqnarray*}
We choose then, up to a normalization factor,
$\eta_{\tau A}(\mathbf{\kappa}) = \left[ \tau\kappa_\perp'+i\kappa_\parallel'\right]$ and
$\eta_{\tau B}(\mathbf{\kappa}) = \pm\abso{\eta_{\tau A}}\, e^{-i\tau\theta}.$
We can see from (\ref{eq:kappa-parallel}) and (\ref{eq:eigenvalues}) that the energies of states with $\kappa_\parallel$ and $-(\kappa_\parallel+2\Delta k_\parallel^c)$ are the same. The energy eigenstate is therefore a linear combination of both:
\begin{eqnarray}
\psi_\tau(\mathbf{r},E_\pm) & = & a_1\; \Phi_\tau(\mathbf{r}, (\kappa_\perp, \kappa_\parallel)) \nonumber \\ 
 & + & a_2\; \Phi_\tau(\mathbf{r}, (\kappa_\perp, -(\kappa_\parallel+2\Delta k_\parallel^c)). 
\end{eqnarray}
From the structure of the lattice in Fig. \ref{fig:general}(a) we see that when the graphene patch is rolled in order to create a zigzag nanotube, the lower CNT edge is formed entirely by $B$ sublattice atoms while the upper edge only by $A$ sublattice atoms. Therefore the wave function on this patch must vanish at the ``missing'' $A$ atoms below the lower edge ($x_\parallel = 0$) and $B$ atoms above the upper edge ($x_\parallel = L$). The conditions for the sublattice components of $\psi_\tau(\mathbf{r},E_\pm)$ are
\begin{subequations}
\label{eq:boundary}
\begin{eqnarray}
& \psi_{\tau A}((x_\perp,0),E_\pm)  \overset{!}{=} 0 \nonumber \\ 
& \rightarrow  a_1 ( \tau\kappa_\perp'+i\kappa_\parallel')
+ a_2 ( \tau\kappa_\perp' -i\kappa_\parallel') 
 = 0,\quad\quad \\[.2cm]
& \psi_{\tau B}((x_\perp,L),E_\pm)  \overset{!}{=} 0 \nonumber \\
& \rightarrow  a_1 \, e^{i\kappa_\parallel' L-i\Delta k_\parallel^c L}
+ a_2 \, e^{-i\kappa_\parallel' L - i\Delta k_\parallel^c L} = 0.
\end{eqnarray}
\end{subequations}
These equations lead to a constraint on the values of $\kappa_\parallel'$,
\begin{equation}
\label{eq:constraint}
\frac{\tau\kappa_\perp' +i \kappa_\parallel'}{\tau\kappa_\perp'-i\kappa_\parallel'} = e^{2i\kappa_\parallel' L}. 
\end{equation} 
Thus the allowed values of $\kappa_\parallel$ depend on $\kappa_\perp'$, and in particular on the Aharonov-Bohm flux $\phi$. The quantity $\kappa_\parallel'$ can be either real or imaginary. If it is real, the wave function describes an {\em extended} state. 
If $\kappa_\parallel'$ is imaginary, then $\kappa_\parallel$ must be complex, with its real part equal to the second term in (\ref{eq:kappa-parallel}). The equation (\ref{eq:constraint}) has then
one trivial ($\kappa_\parallel' = 0$) and two non-trivial solutions. The latter describe {\em evanescent} waves localized near the ends of the nanotube, because the factor $\exp[i(\tau\mathbf{K}+\mathbf{\kappa})\cdot\mathbf{r}_{pi}]$ from (\ref{eq:bloch-waves}) acquires a damping  real part.\\
The regions where $\kappa_\parallel'$ is real or imaginary are determined by the value of $\tau\kappa_\perp'$ (see Fig. \ref{fig:general}(c)). The two localized state solutions exist if
\begin{equation}
\label{eq:localized-exist}
\text{for } K: \; \kappa_\perp'  >  1/L, \quad
\text{for } K': \; \kappa_\perp' < -1/L.
\end{equation}
The spectrum of the CNT is then determined by the value of the magnetic field, which enters into $\kappa_\perp'$ via (\ref{eq:kappa-perp}). In order to calculate the energy levels, the allowed values of $\kappa_\parallel$ must be found from (\ref{eq:constraint}) for each value of $\kappa_\perp'$ separately, as shown in Fig \ref{fig:general}(d) for an (18,0) zigzag CNT.\\
The analytical method described above gives a remarkable agreement with the spectra obtained by the numerical diagonalization of the nanotube Hamiltonian (\ref{eq:hamiltonian-real}), see Figs. \ref{fig:18-0}(a) and \ref{fig:12-9}(a).
The energy of the decaying states tends to 0 with increasing magnetic flux because for $\phi\rightarrow\infty$, $\lvert i \kappa_\parallel'\rvert\rightarrow\kappa_\perp'$ for the $K$ point solutions (see Fig. \ref{fig:general}(d)).
The CNT spectrum may contain localized ($E=0$) states even at $\phi = 0$, as can be seen in Fig. {\ref{fig:18-0}}(a) for an (18,0) CNT.  
If the higher ($\kappa_\perp\neq 0$) subbands lie on the Dirac cone and the condition (\ref{eq:localized-exist}) is fulfilled, then the lowest $\kappa_\parallel$ states in the neighbouring subbands are localized, while the remaining ones have energies in a higher range, appropriate for their subband. Whether the other subbands lie on the Dirac cone depends on the chirality and diameter of the CNT.

\noindent{\em Spin effects.} With spin, the Bloch waves (\ref{eq:bloch-lattice}) become 4-component spinors and both the spin-orbit coupling (SOC) and the Zeeman effect must be considered. They will be treated in detail  elsewhere. Here we just note that SOC can be taken into account by yet another shift of $\kappa_\perp$, while the Zeeman effect splits the energy:
\begin{eqnarray}
\label{eq:so-shift}
\kappa_\perp' &\rightarrow& \kappa_\perp'+ \sigma\,\Delta k_{SO},\quad E_\pm \rightarrow E_\pm + \sigma \mu_B \frac{\phi}{\pi R^2},\quad\\
& & \Delta k_{SO} = \frac{2\delta}{R}\;\left( 1+ \frac{3}{8}\,\frac{V_\sigma-V_\pi}{V_\pi}\right),
\end{eqnarray}
where $\sigma = +1/-1$ for spin parallel/antiparallel to the CNT axis, $\mu_B$ is the Bohr magneton and $\delta$ is a parameter defining the SOC strength. In our calculations we take $\delta \sim 2.8\cdot 10^{-3}$, as e.g. in \cite{jhang:prb2010}. The resulting spectra of a (18,0) and (12,9) CNTs are shown in Figs. \ref{fig:18-0}(c) and \ref{fig:12-9}(b). 

{\em Localization}. Equations (\ref{eq:localized-exist}) define the localization flux $\phi_{loc}$, at which the extended solutions morph into localized states. This threshold flux depends on the spin via (\ref{eq:so-shift}) and using it together with (\ref{eq:kappa-perp}) we obtain
\begin{equation}
\label{eq:philoc-spin}
\phi_{loc} = \tau R\left(\frac{1}{L} -\Delta k_\perp^c\right) \phi_0,\quad
\phi_{loc}^\sigma = \phi_{loc}+\sigma R \Delta k_{SO}\,\phi_0.
\end{equation}
%
%
%
The value of $\phi_{loc}$ depends on the length of the nanotube. For sufficiently long CNTs the spectra are very close to those of the infinite nanotubes.\\
The localization induced by the magnetic field is gradual, in principle allowing the involved states to conduct as long as the two sublattice wavefunctions overlap. The evolution of an eigenstate in increasing magnetic field is shown in Fig. \ref{fig:12-9}(c), through a sequence of plots of the wave function amplitude at each atom, projected onto $x_\parallel$. The apparent continuity of the curves is due to the overlap between close plot points; near the CNT ends the wave function oscillates with the azimuthal angle $\varphi$ and the individual points can be seen clearly. Initially ($\phi=10^{-3}\phi_0$) the state is extended; when the magnetic flux reaches $\phi_{loc}$, its wave function begins to be described by an imaginary solution of (\ref{eq:constraint}). With magnetic field increasing further, the wave function decays exponentially with the distance from the CNT ends, the localization becomes complete and the state ceases to conduct. \\
The above analysis is confirmed by conductance calculations, with the CNT in a setup shown in Fig. \ref{fig:18-0}(b). 
We derive the elastic linear response conductance via the Fisher-Lee formula for the quantum mechanical transmission: $G = \frac{2e^2}{h}\,\text{Tr}\{\mathbf{\Gamma}_L \mathcal{G} \mathbf{\Gamma}_R\mathcal{G}\}$, where $\mathbf{\Gamma}_{L/R} = i(\mathbf{\Sigma}_{L/R}-\mathbf{\Sigma}_{L/R})$,  $\mathbf{\Sigma}_{L/R}$ is the self energy of the left or right lead respectively, and $\mathcal{G}$ is the Green function of the central region dressed by the electrodes. For simulating bulk metal electrodes we consider wide band leads, i.e. $\mathbf{\Sigma}_{WB}(E) = -i\,\text{Im}\mathbf{\Sigma}(E_F)$. The results shown in Figs. \ref{fig:18-0}(d) and \ref{fig:12-9}(d) were obtained with $\mathbf{\Sigma}_{WB} = -i\,0.22$eV. In both we see a gradual drop of the conductance of the highest valence and lowest conduction spin states, as they become localized in the increasing magnetic field. The ``native'' end states of the (18,0) CNT, localized also at $\phi = 0$, can be seen in the spectrum in Fig. \ref{fig:18-0}(c), but don't contribute to the conductance, as we expect. The good matching of analytical spectra of isolated CNTs and the conductance peaks (compare Figs. \ref{fig:18-0}(c),(d) and \ref{fig:12-9}(b),(d)) implies that even with rather strong coupling the CNT  is sufficiently distinct from the leads for the transport to be determined by the spectrum of an isolated nanotube.\\
%
%
The magnetic field $B_{loc}$ corresponding to $\phi_{loc}$ depends on the nanotube length and radius. For our choice of $V_\pi$ and $V_\sigma$, $B_{loc}$ of CNTs with $R=7$ {\AA} and $L = 40$ nm ranges from 50 T ($\theta\approx 30^\circ$) to 85 T ($\theta = 0^\circ$). However, for the same nanotubes with $L = 500$ nm the value of $B_{loc}$ drops to 4-42 T. Hence the localization induced by the magnetic field might be detected in currently accessible transport experiments or by STM spectroscopy revealing localized states at the CNT ends. 
Moreover, the localization induced by a magnetic flux appears to be a chirality-independent phenomenon, to which only armchair CNTs are immune.

\acknowledgements{
The authors acknowledge the support of the Deutsche Forschungsgemeinschaft under the GRK grant 1570.
}


\end{document}